\documentclass[9pt,journal,twoside,web]{ieeecolor}
\usepackage{generic}
\usepackage{cite}
\usepackage{amsmath,amssymb,amsfonts}
\usepackage{graphicx}
\usepackage{textcomp}
\usepackage{hyperref}
\usepackage{xcolor}
\hypersetup{
    colorlinks,
    linkcolor={red!50!black},
    citecolor={blue!50!black},
    urlcolor={blue!80!black}
}

\usepackage{booktabs}
\usepackage{multirow}
\usepackage{adjustbox}
\usepackage{threeparttable}

\usepackage{float}
\usepackage{caption}
\usepackage{subcaption}

\usepackage[vlined,ruled]{algorithm2e}

\def\BibTeX{{\rm B\kern-.05em{\sc i\kern-.025em b}\kern-.08em
    T\kern-.1667em\lower.7ex\hbox{E}\kern-.125emX}}
\begin{document}
\title{PseudoCell: Hard Negative Mining as Pseudo Labeling for Deep Learning-Based Centroblast Cell Detection}
\author{Narongrid Seesawad, 
Piyalitt Ittichaiwong, 
Thapanun Sudhawiyangkul,
Phattarapong Sawangjai,
Peti Thuwajit,
Paisarn Boonsakan,
Supasan Sripodok, 
Kanyakorn Veerakanjana,
Phoomraphee Luenam,
Komgrid Charngkaew, 
Ananya Pongpaibul, 
Napat Angkathunyakul, 
Narit Hnoohom,
Sumeth Yuenyong,
Chanitra Thuwajit, and
Theerawit Wilaiprasitporn \IEEEmembership{Senior Member, IEEE}
\thanks{This work was supported by grants from New Discovery \& Frontier Research Grant, Mahidol University.
\textit{(Corresponding authors: Piyalitt Ittichaiwong, Thapanun Sudhawiyangkul, Chanitra Thuwajit, and Theerawit Wilaiprasitporn.)}}
\thanks{N. Seesawad, T. Sudhawiyangkul, P. Sawangjai, P. Luenam, and T. Wilaiprasitporn are with the Bio-inspired Robotics and Neural Engineering (BRAIN) Lab, School of Information Science and Technology (IST), Vidyasirimedhi Institute of Science \& Technology (VISTEC), Rayong, Thailand (theerawit.w@vistec.ac.th)}
\thanks{P. Ittichaiwong, and K. Veerakanjana are with the Siriraj Informatics and Data Innovation Center, Faculty of Medicine Siriraj Hospital, Mahidol University, Bangkok, Thailand}
\thanks{P. Thuwajit, and C. Thuwajit are with the Department of Immunology, Faculty of Medicine Siriraj Hospital, Mahidol University, Bangkok, Thailand}
\thanks{P. Boonsakan is with the Department of Pathology, Faculty of Medicine Ramathibodi Hospital, Mahidol University, Bangkok, Thailand}
\thanks{S. Sripodok, K. Charngkaew, A. Pongpaibul, and N. Angkathunyakul are with the Department of Pathology, Faculty of Medicine Siriraj Hospital, Mahidol University, Bangkok, Thailand}
\thanks{N. Hnoohom, and S. Yuenyong are with the Department of Computer Engineering, Faculty of Engineering, Mahidol University, Nakhon Pathom, Thailand}
}

\maketitle

\begin{abstract}
Patch classification models based on deep learning have been utilized in whole-slide images (WSI) of H\&E-stained tissue samples to assist pathologists in grading follicular lymphoma patients. However, these approaches still require pathologists to manually identify centroblast cells and provide refined labels for optimal performance. To address this, we propose \textit{PseudoCell}, an object detection framework to automate centroblast detection in WSI (source code is available at \url{https://github.com/IoBT-VISTEC/PseudoCell.git} [Note: The code will be public after the manuscript is accepted]). This framework incorporates centroblast labels from pathologists and combines them with pseudo-negative labels obtained from undersampled false-positive predictions using the cell's morphological features. By employing \textit{PseudoCell}, pathologists' workload can be reduced as it accurately narrows down the areas requiring their attention during examining tissue. Depending on the confidence threshold, \textit{PseudoCell} can eliminate 58.18–99.35\% of non-centroblasts tissue areas on WSI. This study presents a practical centroblast prescreening method that does not require pathologists' refined labels for improvement. Detailed guidance on the practical implementation of \textit{PseudoCell} is provided in the discussion section.
\end{abstract}


\section{Introduction}
\label{sec:introduction}
Follicular lymphoma (FL) is the second most prevalent lymphoid malignancy in Western and Asian countries. It is responsible for 5-35\% of non-Hodgkin lymphoma (NHL) \cite{suzumiya2018current, swerdlow20162016, intragumtornchai2018non}. Most FL carries the translocation t(14;18), which causes the overexpression of the BCL-2 protein. FL patients usually present with lymphadenopathy, infrequent B-symptoms, systemic fever symptoms, night sweats, and weight loss. The progression of a disease can be predicted using a combination of clinical and laboratory findings, as well as the histopathological grade of the disease \cite{mozas2021past}
 
\begin{figure}[t]
    \centering
    \includegraphics[width=\columnwidth]{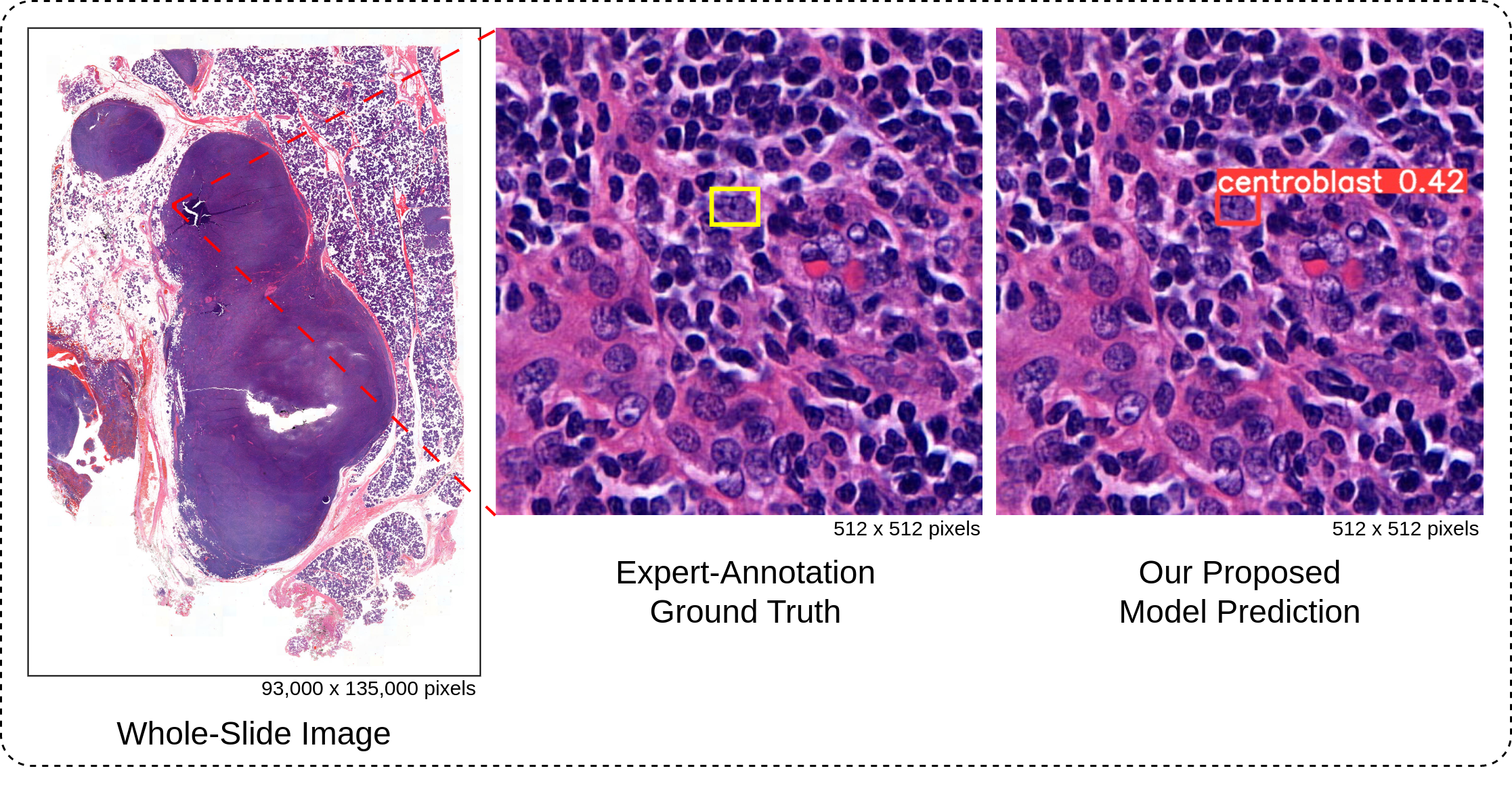}
    \caption{\textbf{Automated detection of centroblast cells in whole-slide images.} The tiny red square within WSI represents the patch image. Experts spent much time examining WSI under a microscope, patch by patch, to identify CB. In contrast, our proposed model can immediately identify CB with a confidence score.}
    \label{fig:intro}
\end{figure}

Currently, the WHO classification system is the standard for FL grading. The grading system is identified by the number of large neoplastic cells in a tissue sample, known as centroblast cells (CB). In the conventional approach, pathologists rely on manual counting of these CB in tissue samples stained with hematoxylin and eosin (H\&E) under a microscope. However, this procedure is time-consuming, laborious (\autoref{fig:intro}), and subjective because of differences in the expert's experience. This results in high inter- and intra-observer variability, 61-73\% \cite{lozanski2013inter}. The high variability is causing the data to be susceptible to sampling bias, hard to reproduce, and directly impacting patients' clinical management since there is a lack of consensus among pathologists \cite{lozanski2013inter}. Therefore, enhancing the precision, reliability, and reproducibility of histological grading is highly significant.

Numerous studies have proposed automated methods to localize and classify FL by using whole-slide images (WSI), scanned images from the tissue samples, aiming to facilitate the work of pathologists. The techniques can be categorized into two groups: 1) machine learning (ML)-based approaches with human-engineered features \cite{belkacem2009extraction, cheng2010identifying, michail2014detection, dimitropoulos2014using, sertel2010computer, dimitropoulos2017automated} and 2) deep learning (DL)-based approaches \cite{liu2017detecting, lu2020deep, somaratne2019improving, syrykh2020accurate}. Still, the first approach is feeding handcrafted features into ML, which tends to be overfitting, has high false-positive (FP) prediction, and is difficult to generalize \cite{cheng2010identifying, michail2014detection, dimitropoulos2014using, sertel2010computer}. Because the performance of the model heavily depends on the combination of features they use. Therefore, many studies have used DL-based methods to eliminate the need for hand-engineer features and extract essential features from the training dataset.
 
DL-based models, especially Convolutional Neural Network (CNN), have been recently applied to detect and classify lymph nodes on H\&E-stained WSIs. To detect lymphocytes in breast cancer (BC), Liu et al. \cite{liu2017detecting} addressed the tumor class imbalance problem by applying random sampling and data augmentation on patches (i.e., cropped images from WSI) before training the InceptionV3 \cite{szegedy2015going}. Their method had the best sensitivity on the Camelyon16 dataset. To obtain a robust model in a new cohort, Lu et al. \cite{lu2020deep} proposed an automatic pipeline employing cascade training on U-Net \cite{ronneberger2015u}. The pipeline is an iterative training process where the model is finetuned on the new cohort using its predicted lymphocyte mask, which was evaluated and refined by pathologists. Two iterations of cascade training were repeated to produce a model with an F1-score of 0.927. However, obtaining the refined mask also increases the workload of experts, contradicting the intended reduction in pathologists' workload.

Comparing FL and BC, most DL studies in FL WSIs have focused on patch-level classification (i.e., whether patches contain CB), resulting in lower interpretability to grade FL. In 2019, Somaratne et al. \cite{somaratne2019improving} developed the one-class training approach to minimize the generalization gap between two FL datasets by combining several images from the target set into the training set. Then apply transfer learning to AlexNet \cite{krizhevsky2017imagenet} on the new training set. The transfer learning model improves patch classification accuracy at the patch level by 12\% over the model that trains from scratch. In 2020, Syrykh et al. \cite{syrykh2020accurate} used a CNN-based model to differentiate between FL and follicular hyperplasia (FH) at four different resolutions in histopathology slides. It resulted that the model with the highest resolution achieved accurate patch-level classification. However, this study also showed that the performance of the DL-based approaches is sensitive to the pre-processing of histopathology slides, as evidenced by the drop in the area under the curve (AUC) from 0.92-0.99 (internal dataset) to 0.63-0.69 (external dataset). The DL approach still needs stain normalization (SN) in pre-processing phase.

\section{Motivation and contribution}
According to the limitations mentioned above: (1) Deep Learning (DL) is sensitive to the variation of stain color in WSIs; (2) refined labels from experts are required to improve the model during training; and (3) class imbalance between the CB and non-CB classes appears in the dataset. These limitations restricted DL's improvement on FL WSIs to cell-level prediction. 

To overcome these limitations, we proposed a framework called \textit{PseudoCell} to explore the feasibility of DL-based object detection models on CB detection tasks. We aim to use the state-of-the-art object detection model, YOLOv8 \cite{Jocher_YOLO_by_Ultralytics_2023}, as our backbone model. Firstly, we compare the consistency of two Stain Normalization (SN) methods on our dataset to prevent the effect of color variation from WSI. Secondly, the need for expert refined labels during training will be imitated through the hard negative mining technique (HNM) \cite{shrivastava2016training}, i.e., retrieving false-positive (FP) predictions from the trained model, afterward incorporating them into the training set as pseudo-negative labels (non-CB class), and training a new model. Since the number of pseudo-negative labels is higher than the number of CB labels from pathologists, the imbalance class issue must be addressed before incorporating pseudo-negative labels. Thirdly, three distinct undersampling approaches were explored to mitigate the class imbalance issue before incorporating pseudo-negative labels into the training set. 

To our best knowledge, HNM was initially introduced in the field of computer vision and has yet to be utilized in the context of digital histopathological image recognition. While previous work on cancer cells seeking refined labels from experts to enhance the model, we instead attempted to imitate it through the HNM. This framework allows us to improve the model autonomously without relying on additional work from pathologists. Therefore, the comparison between different HNM approaches was mainly investigated. 

Lastly, we have provided a practical guideline based on high-power field selection and CB identification in WSI for the practical application of our \textit{PseudoCell} as a pre-screening tool for FL patients. Integrating this framework with histopathological workflow can reduce experts' workload by narrowing down the region experts focus on while examining the tissue. Other potential real-world applications (such as quality control, training, and education tools) are also discussed for the benefit of human-machine collaboration.

\section{Methods}
\subsection{Data collection}
This study included 75,245 patches (512x512 pixels) of Follicular Lymphoma (FL) admitted for treatment at the Faculty of Medicine Siriraj Hospital between 2016 and 2020. No significant correlation between clinicopathological parameters was observed (data not shown). The Siriraj Institutional Review Board (SIRB) (COA no. Si973/2020) has approved the procedures for obtaining and using tissue. Formalin-fixed paraffin-embedded (FFPE) tissue samples with a thickness of 3-5 microns were prepared for automated hematoxylin and eosin (H\&E) staining and scanned at a resolution of 0.12 microns per pixel using a 3Dhistech Panoramic 1000 microscope with a 40x objective lens. The resulting images were saved in NRXS format.
 
From a total of 75,245 patches, 1203 patches contain Centroblast (CB) cells, and 3045 patches without CB were selected and annotated by a consensus of two doctors (one of them is a pathologist). The annotation is manually drawn around CB as a bounding box (bbox), \autoref{fig:fw}(a).

\subsection{The Proposed Framework}
\label{sec:proposed_framework}
Based on the challenge of CB cell detection, we proposed a framework in \autoref{fig:fw}(b)-\autoref{fig:fw}(d) that gives reproducible cell-level predictions. Our proposed framework comprises three parts: 1) Train original model, 2) Hard Negative Mining pipeline, and 3) Train model with negative pseudo label.

\subsubsection{Train original model}
As shown in \autoref{fig:fw}(b), three steps comprise this part to obtain an one-class dataset and a CB detection model: 1.1) Stain normalization selection; 1.2) Data preprocessing; 1.3) Model training.

\newcounter{paranum}[subsubsection]
\newcommand{\PP}{\textbf{1.\refstepcounter{paranum}\theparanum)} \textit}

\PP{\textbf{Stain normalization selection:}}
Even though our WSIs came from the same lab and scanner, the WSIs still have the variation in stain colors. So Stain normalization is applied to our preprocessing step. 

Stain normalization (SN) is the color distribution transformation from a source image $I$ into a target image $I^{\prime}$. The transformation can be described through the operation $I^{\prime}=f(I, \theta)$ where $\theta$ is a collection of parameters derived from the template image, and $f$ is the function that maps the visual appearance of a given image $I$ to the template image. Generally, $\theta$ is designed to capture the color information of the primary stain components (e.g., hematoxylin and eosin). Consequently, stain-normalized images will have a color distribution similar to the template image \cite{ciompi2017importance}.

In this work, we consider two state-of-the-art SN methods:
    \begin{itemize}
        \item \textbf{Structure Preserving Color Normalization (SPCN):} Vahadane \textit{et al.} proposed in \cite{vahadane2016structure}, which tackles the stain separation problem with the assumption that stain density is non-negative, and the color basis is sparse. The sparseness constraint reduces the solution space of the color decomposition problem. Then, the color basis of a source image is replaced with those from a template image while maintaining its original stain concentrations. 
        
        \item \textbf{Deep convolutional Gaussian mixture models (DCGMM):} Zanjani \textit{et al.} proposed in \cite{zanjani2018histopathology}. This method first converts the source image into the HSD color system. Then fits a GMM to the color distribution individually per tissue class. To train the DCGMM, \textit{E-step} and \textit{M-step} of the EM-algorithm are replaced by gradient descent and the back-propagation algorithm. The advantage of this approach is that it does not need any assumptions about the H\&E image content.
    \end{itemize}
We conduct an experiment, detailed in Section \ref{sec:exp1_setup}, to compare and select the most appropriate SN method for our dataset (i.e., one that produces processed images with low color variation and minimal background error).

\PP{\textbf{Data preprocessing:}}
Due to the considerable human errors during annotation, label cleaning was necessary before feeding data into the model. The two most prevalent errors in our dataset are 1) bbox annotations with zero areas and 2) repeated bbox annotations on a single CB cell. Since the annotator may have accidentally generated a bbox with zero areas by clicking the mouse, we removed all bbox annotations with zero areas from our dataset. Regarding the second error, we first calculated the center of each bbox and then retrieved the groups of bounding boxes whose center-to-center distance is within a constant. If bbox annotations share the same CB cell, we select the bbox that best fits the cell based on manual inspection of each bbox group.  

Then apply the stain normalization method from the previous experiment to the annotated positive patches in order to standardize the color variation on our dataset. Lastly, 80\%, 10\%, and 10\% of the normalized positive patches were separated into train, validation, and test sets to create dataset $D_{1}$.

\PP{\textbf{Model training:}}
Before feeding the training set into the model, five augmentation methods (flip up-down, flip left-right, rotate 90 degrees, rotate 180 degrees, and rotate 270 degrees) were applied to the training set.

YOLOv8 (architecture: \textit{X6}) was trained and validated using 10-fold cross-validation on the augmented dataset $D_1$ with default hyperparameter configuration. Stochastic gradient descent (SGD) was applied to reduce cross-entropy loss. The model was trained for a maximum of 500 epochs, with early stopping to terminate training when the validation loss stopped improving. Ultimately. We eventually obtained the original \textit{ori} model.

\subsubsection{Hard Negative Mining Pipeline} 
\label{sec:hnm_pipe}
In histopathological image recognition, pathologists typically annotate only target cells (i.e., CB cells) and leave other cells unannotated to minimize the annotation cost. It causes DL-based models to typically perform poorly due to many false-positive (FP) predictions. 

We hypothesize that distinguishing CB cells from other cells that look like CB cells (non-CB cells) is the key to improving the model. One approach is incrementing the non-CB labels as a new class in the dataset. In practice, we retrieve the FP bbox (i.e., non-CB annotation) from the \textit{ori} model inference on the training set and add them to the training set as a new class. As shown in \autoref{fig:fw}(c), the following three steps were employed to generate a dataset with pseudo-negative labels: 2.1) Retrieve FP predictions; 2.2) Undersample; and 2.3) Combine the non-CB class with the training set.

\newcounter{paranumPPP}[subsubsection]
\newcommand{\PPP}{\textbf{2.\refstepcounter{paranumPPP}\theparanumPPP)} \textit}

\PPP{\textbf{Retrieve FP predictions:}}
To obtain FP predictions, we let \textit{ori} infer the training set in each fold using a confidence threshold of 0.001 to ensure that the model predicts all negatives. From the training set, the long side of CB bounding box (bbox) is smaller than 100 pixels, so we filter out the FP bbox with a box side greater than 100 pixels.

\PPP{\textbf{Undersample:}}
Since the number of FP predictions is still greater than that of CB, directly adding negatives into the training set will result in an imbalance class problem. We consider two undersampling strategies to prevent the imbalance issue: Random undersampling and Neighborhood-based Recursive search undersampling (NB-REC) \cite{vuttipittayamongkol2020neighbourhood}.
    \begin{itemize}
        \item \textbf{Random undersampling} is a popular non-heuristic technique due to its simplicity of application. Despite its simplicity, there is a significant disadvantage that must be considered. Given that balanced class distribution is a stopping criterion, random undersampling may eliminate potentially useful samples in order to achieve this balance \cite{lopez2013insight}.
        \item \textbf{NB-Rec} eliminates the majority class sample, which may overlap with minority class. As described in Algorithm \ref{algo:nb_rec}, the majority sample is considered overlapping when it is in the neighborhood of more than one minority sample. Since the NB-Rec uses K-Nearest Neighbor (KNN), we must search for $k$ prior to execution in order to produce a number of negatives approximately equal to CB. 
    \end{itemize}

Given that the coordinate is necessary for NB-Rec undersampling, we must extract features from both the ground truth bbox and the FP bbox. The width and height of the bbox were extracted directly from the samples. Six morphological features, described in \cite{vrabac2021dlbcl}, were calculated using a binary image of each cell in bbox segmented by a trained HoverNet model based on the \textit{PanNuke} architecture provided in \cite{graham2019hover}.

As depicted in \autoref{fig:fw}(c) by the red, blue, and purple paths, we obtain three sets of undersampled FP predictions: (1) the set from random undersampling, (2) the set from applied NB-Rec undersampling to bbox width and height, and (3) the set from NB-Rec undersampling applied to the first- and second-principal components of bbox width, bbox height, and six morphological features using the Principal Component Analysis (PCA) method. 

\PPP{\textbf{Combine the non-CB class with the training set:}}
At each path from the previous step, the undersampled FP predictions are added to the training set of dataset $D_{1}$. Similar to $D_1$, the new three independent datasets with pseudo-negative labels contain two classes (CB and non-CB) in the training set, but only one class (CB) in the validating and testing sets.

\makeatletter
\renewcommand{\@algocf@capt@plain}{above}
\makeatother
\begin{algorithm}[t]
\caption{NB-Rec}\label{algo:nb_rec}

\textbf{Input:} training set, $k$ \\
\textbf{Output:} undersampled training set

\SetKwBlock{Beginn}{beginn}{ende}
\DontPrintSemicolon
\Begin{

    $T \gets$ training set\;
    $T_{pos} \gets$ positive instances in $T$\;
    $A \gets $ frequency table\;
    \For{$x \in T_{pos}$}{
        $NN \gets k$ nearest neighbours\;
        $NN_{neg} \gets $ negative members of $NN$\;

        \For{$y \in NN_{neg}$}{
            $A_{y}.freq \gets A_y.freq + 1$\;
        }
    }    
    \For{$x \in A.instance$}{
        \If{$A_x.freq > 1$}{
            $X \gets X \cup {x}$
        }
    } 
    \;
    $A^\prime \gets $ frequency table\;
    \For{$x_1 \in X$}{
        $NN_2 \gets k$ nearest neighbours\;
        $NN_{2_{neg}} \gets $ negative members of $NN_2$\;
        
        \For{$y \in NN_{2_{neg}}$}{
            $A^{\prime}_{y}.freq \gets A^\prime_y.freq + 1$\;
        }
    }
    \For{$x_2 \in A^{\prime}.instance$}{
        \If{$A^{\prime}_{x_2}.freq > 1$}{
            $X_2 \gets X_2 \cup {x_2}$
        }
    }
    $T^\prime \gets T \setminus (X \cup X_2)$\;
    \Return $T^\prime$
}
\end{algorithm}

\subsubsection{Train model with pseudo-negative label} 
\begin{figure*}[!h]
    \centering
    \includegraphics[width=0.85\textwidth]{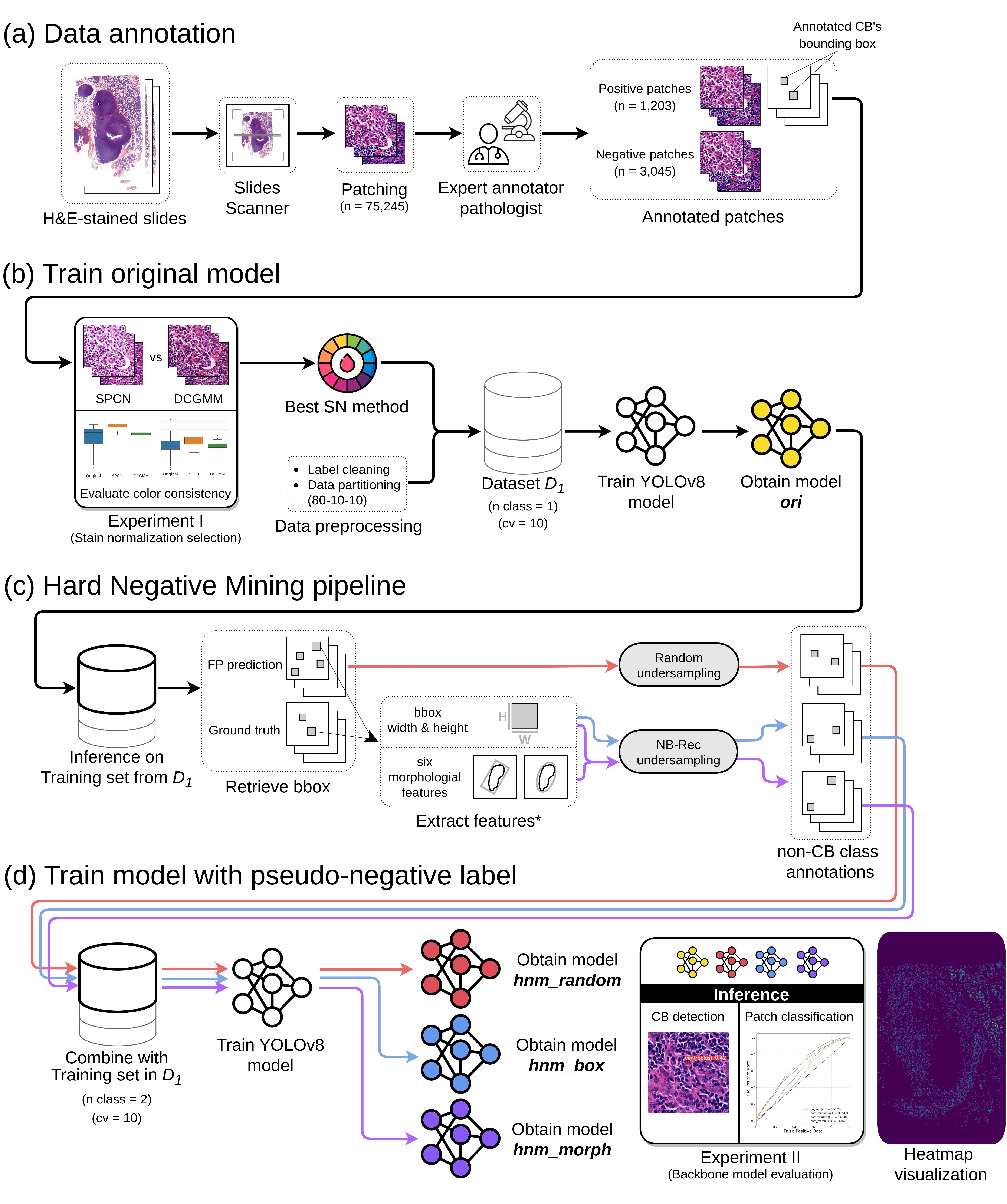}
    \caption{\textbf{Overview of the \textit{PseudoCell} framework.} \\
    \textbf{(a) Data annotation:} H\&E-stained slides were scanned and patched into 512x512 pixels. An expert pathologist annotated centroblast (CB) cells on selected patches by locating a rectangle bound to each CB while another expert pathologist reviewed the annotations. \\
    \textbf{(b) Train original model:}  The CB-annotated patches were cleaned and then normalized the color using the best stain normalization method from \textit{Experiment I}. Then, 80\%, 10\%, and 10\% of the data were separated into train, validation, and test sets to create dataset $D_1$. Five methods augmented the training set before being fed into YOLOv8. YOLOv8 was trained and validated in a 10-fold cross-validation manner to generate \textbf{\textit{ori}} model. \\
    \textbf{(c) Hard Negative Mining pipeline:} The \textit{ori} model was then applied to infer the training set in order to retrieve false-positive (FP) samples. We then employed three undersampling strategies (red, blue, and purple paths) to avoid the imbalance class issue. \textbf{*:} the six morphological features \cite{vrabac2021dlbcl} were calculated using a binary image segmented by a trained HoverNet model \cite{graham2019hover}. \\
    \textbf{(d) Train model with pseudo-negative label:} The undersampled FP samples from each path were combined with the training set of dataset $D_1$ as a new class. Consequently, each path had its own training set of two classes with identical validation and test sets from $D_1$. YOLOv8 was trained and validated with similar manner as \textit{ori} model to obtain \textbf{\textit{hnm\_random}}, \textbf{\textit{hnm\_box}}, and \textbf{\textit{hnm\_morph}} models. Finally, all models were compared in \textit{Experiment II}, and the best approach was applied to visualize the heatmap of WSI.}
    \label{fig:fw}
\end{figure*}

As shown in \autoref{fig:fw}(d), similar to the part 1) \textit{Training original model}, we use the same setup \textit{model training step} on the dataset of red, blue, and purple paths to get model \textit{hnm\_random}, \textit{hnm\_box}, and \textit{hnm\_morph}, respectively.

\subsection{Evaluation metrics}
\subsubsection{Stain normalization evaluation}
\label{sec:eval_sn}
In Experiment I, we compare two stain normalizations (i.e., SPCN and DCGMM) in intensity, hue, and color error. Normalized Median Intensity (NMI) \cite{nyul2000new}\cite{bejnordi2015stain} is a popular metric that quantifies the intensity variation of an image population, especially the color constancy of the nuclei. NMI is defined as \autoref{eq:nmi} where $A(i)$ is the average values of Red, Green, and Blue channels at the pixel $i$ in RGB image $I$, and $P_{95}$ is the 95th percentile. 

 \begin{equation}
 \label{eq:nmi}
 NMI(I) = \frac{Median_{i \in I}\{A(i)\}}{P_{95}\{A(i)\}}
 \end{equation}

Normalized Median Hue (NMH) \cite{pontalba2019assessing} is similar to NMI but looks at the consistency of hue and is defined by replacing the $A(i)$ with $H(i)$ which is the value of hue-channel at the pixel $i$ in HSV image $I$.

The standard deviation (SD) and the coefficient of variation (CV)—standard deviation divided by mean—of NMI and NMH were computed to indicate the relative dispersion of measures around the mean of each image population. The lower CV indicates a lower dispersion.

To measure the error in the background of processed images, Absolute Mean Color Error (AMCE) \cite{roy2019novel} is applied on $l\alpha\beta$ (decorrelated) color space for both $\alpha$ and $\beta$ channels, which are given in \autoref{eq:amce_a} and \autoref{eq:amce_b}, respectively. Where $\mu$ is the local mean, $\alpha_{i}(tar)$ is the value of target image at local window $i$ in $\alpha$ channel, $\alpha_{i}(proc)$ is the value of processed image at local window $i$ in $\alpha$ channel, $W$ is the total number of windows, and $\beta$ channel for $\beta_{i}()$.

 \begin{equation}
 \label{eq:amce_a}
 AMCE_{\alpha} = \left|\frac{1}{W}\sum_{i=1}^{W}\mu(\alpha_{i}(tar)) - \frac{1}{W}\sum_{i=1}^{W}\mu(\alpha_{i}(proc))\right|
 \end{equation}
  
 \begin{equation}
 \label{eq:amce_b}
 AMCE_{\beta} = \left|\frac{1}{W}\sum_{i=1}^{W}\mu(\beta_{i}(tar)) - \frac{1}{W}\sum_{i=1}^{W}\mu(\beta_{i}(proc))\right|
 \end{equation}

\subsubsection{Object detection evaluation}
\label{sec:eval_obj}
In Experiment II, the performance of models is evaluated with a certain Intersection over Union (IoU) threshold. IoU can be denoted as follows: 

 \begin{equation}
 \label{eq:iou}
 IoU_{i} = \frac{P_{i} \cap G_{i}}{P_{i} \cup G_{i}}
 \end{equation}

Where $P_{i}$ is the $i$ bounding box (bbox) predicted by the model and $G_{i}$ is the corresponding bbox in the ground truth. If the IoU of a predicted bbox is greater than a 0.5 IoU threshold, the predicted bbox is classified as a true positive (TP). Otherwise, the bbox is classified as a false positive (FP). If the model does not detect the region in the ground truth, the ground truth is classified as a false negative (FN). It should be noted that if more than one predicted bbox matches the same reference bbox, the predicted bbox with the highest IoU is chosen as the TP, and the others are excluded from validation. 

These three elements (i.e., TP, FP, and FN) allowed us to determine precision (P) and recall (R). The average precision (AP) depicts the trade-off between precision and recall at different thresholds, defined as \autoref{eq:ap}, which is the area under the precision-recall curve at different thresholds. The mean average precision (mAP) is the average of each class-specific AP score. Since all models in this work perform single-class prediction, the AP and mAP are equivalent.

 \begin{equation}
 \label{eq:ap}
  AP = \int_{0}^{1} P(R) \,dR 
 \end{equation}

\subsection{Experiment setup:}
\label{sec:exp_setup}
All experiments were performed with an NVIDIA Tesla V100-SXM2 graphic card. 
\subsubsection{Experiment I: Stain Normalization Selection}
\label{sec:exp1_setup}
This experiment compares the color consistency of our dataset after applying stain normalization methods (SPCN and DCGMM). We experiment with a template image that an expert prefers from all patches. Then apply both SN methods to the remaining images. The normalized images were evaluated using metrics in Section \ref{sec:eval_sn}. The best SN method will be used in the pre-processing phase of this work.

\subsubsection{Experiment II: Backbone model evaluation}
\label{sec:exp2_setup}
This experiment aims to compare the performance of models from different training approaches (i.e., conventional and HNM approaches) on both object detection and image classification tasks. We use the training pipeline described in Section \ref{sec:proposed_framework} to obtain four models: 
\begin{itemize}
	\item \textbf{Original (\textit{ori}) model}: Conventional object detection approach with one class annotation.
 
	\item \textbf{Model trained with random HNM (\textit{hnm\_random})}: Randomly add FP samples, from \textit{ori} prediction on the training set, into the training set as a new class. Then trains the model with the same setup as \textit{ori}. 
 
	\item \textbf{Model trained with HNM of bbox features (\textit{hnm\_box})}: Instead of randomly sampling, this approach undersamples the FP samples using NB-Rec on the width and height of the FP bounding box, then adds them into the training set. 
 
	\item \textbf{Model trained with HNM of morphological features (\textit{hnm\_morph})}: Similar to \textit{hnm\_box} but using NB-Rec on first- and second-principal components from six morphological features and width and height of FP bounding box.  
\end{itemize}
We use metrics from Section \ref{sec:eval_obj} to evaluate the performance models on object-level prediction. 

For the image classification task, the object-level prediction was mapped into the image classification using the following criteria: \textit{"For any patch, if a CB prediction exists in cell-level prediction, the patch is classified as a positive patch. If not, the patch was classified as a negative patch."}  Since the test set of each fold contains 120-121 images with CB (i.e., positive images), we evaluate model performance on the test set of each fold with the additional negative image from our database by randomly selecting.

\section{Results}

\subsection{Experimental Results}
\subsubsection{Experiment I: Stain normalization Selection}
\label{sec:result_exp1}
\begin{figure}[t]
    \centering
    \includegraphics[width=\columnwidth]{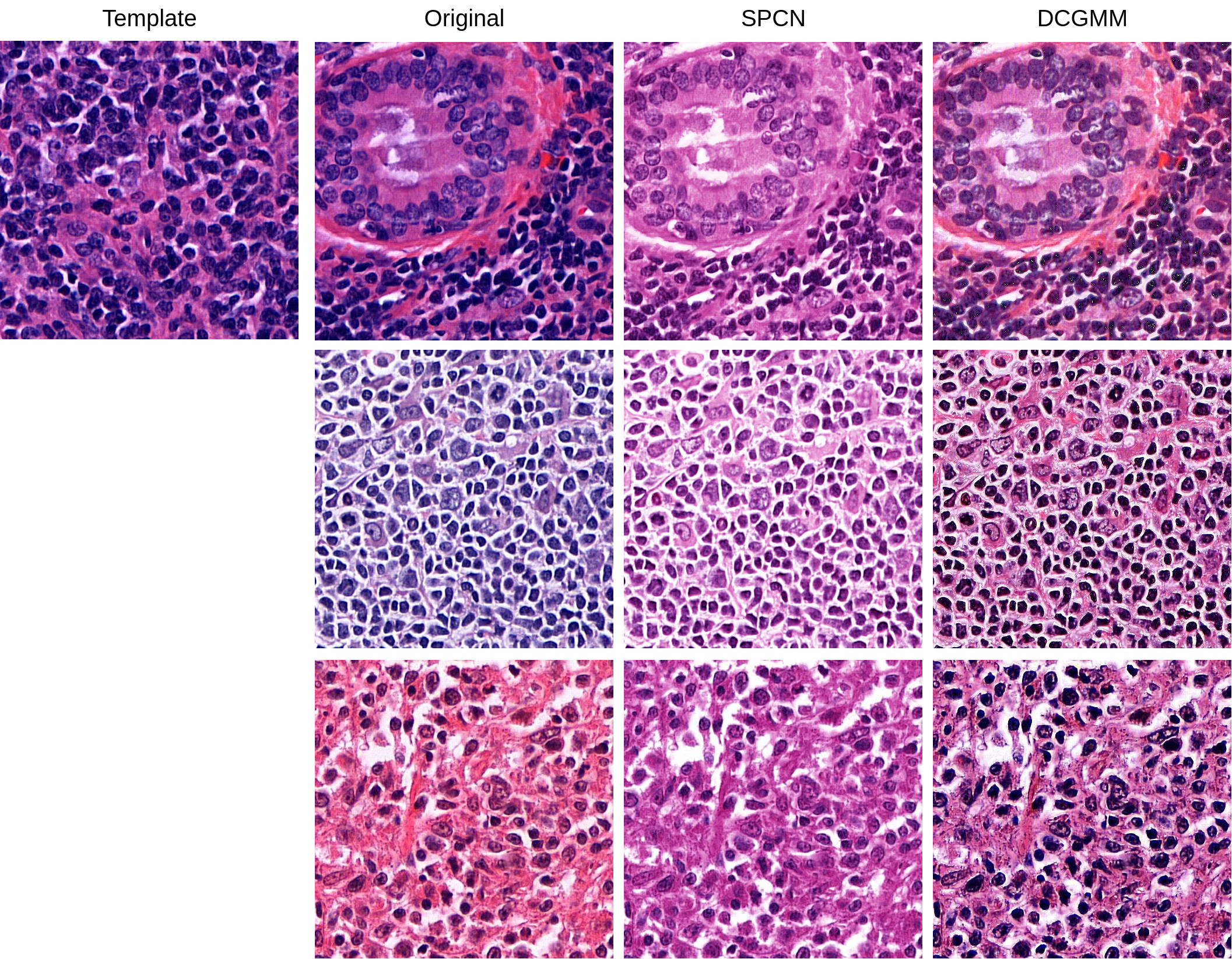}
    \caption{\textbf{Illustration of the performance of different stain normalization methods.} The top-left image is the template image. The next column is the images sampled from the original images, followed by the results of normalization using SPCN and DCGMM, respectively.}
    \label{fig:compare_sn}
\end{figure}

\begin{figure}
     \centering
     \begin{subfigure}[b]{0.49\columnwidth}
         \centering
         \includegraphics[width=\textwidth]{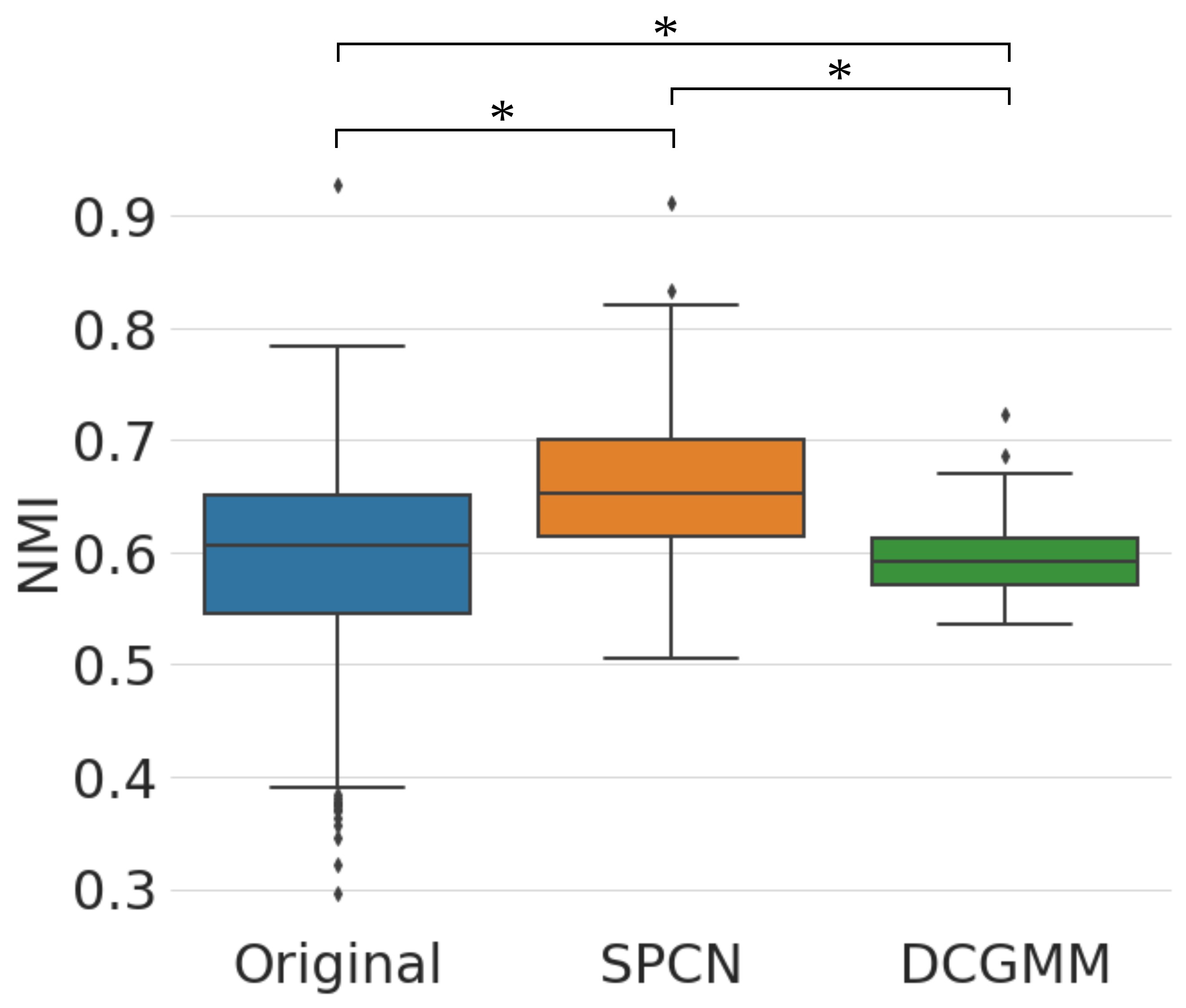}
         \label{fig:nmi}
     \end{subfigure}
     \hfill
     \begin{subfigure}[b]{0.49\columnwidth}
         \centering
         \includegraphics[width=\textwidth]{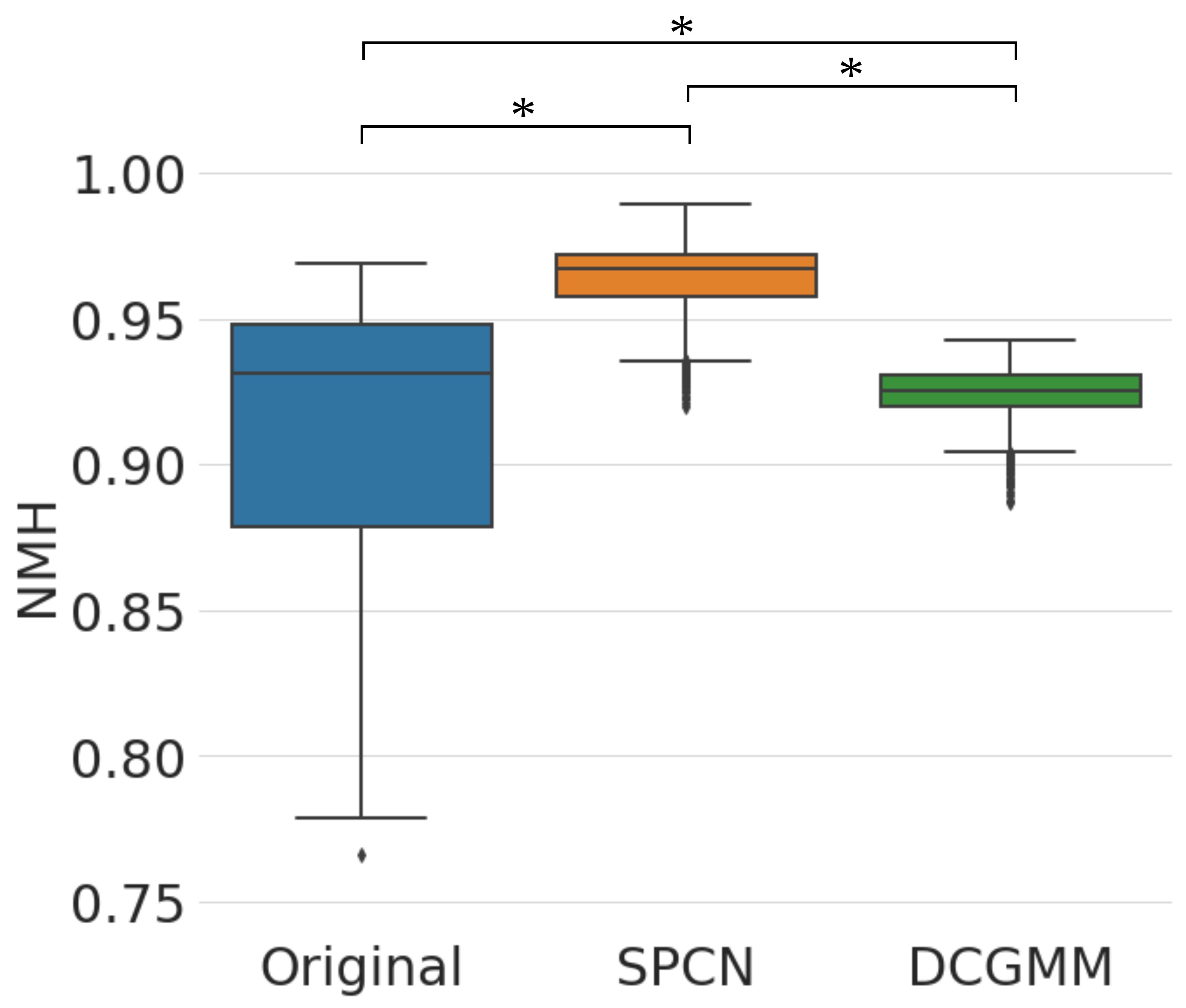}
         \label{fig:nmh}
     \end{subfigure}
    
    \hfill
    \begin{subfigure}[b]{0.49\columnwidth}
         \centering
         \includegraphics[width=\textwidth]{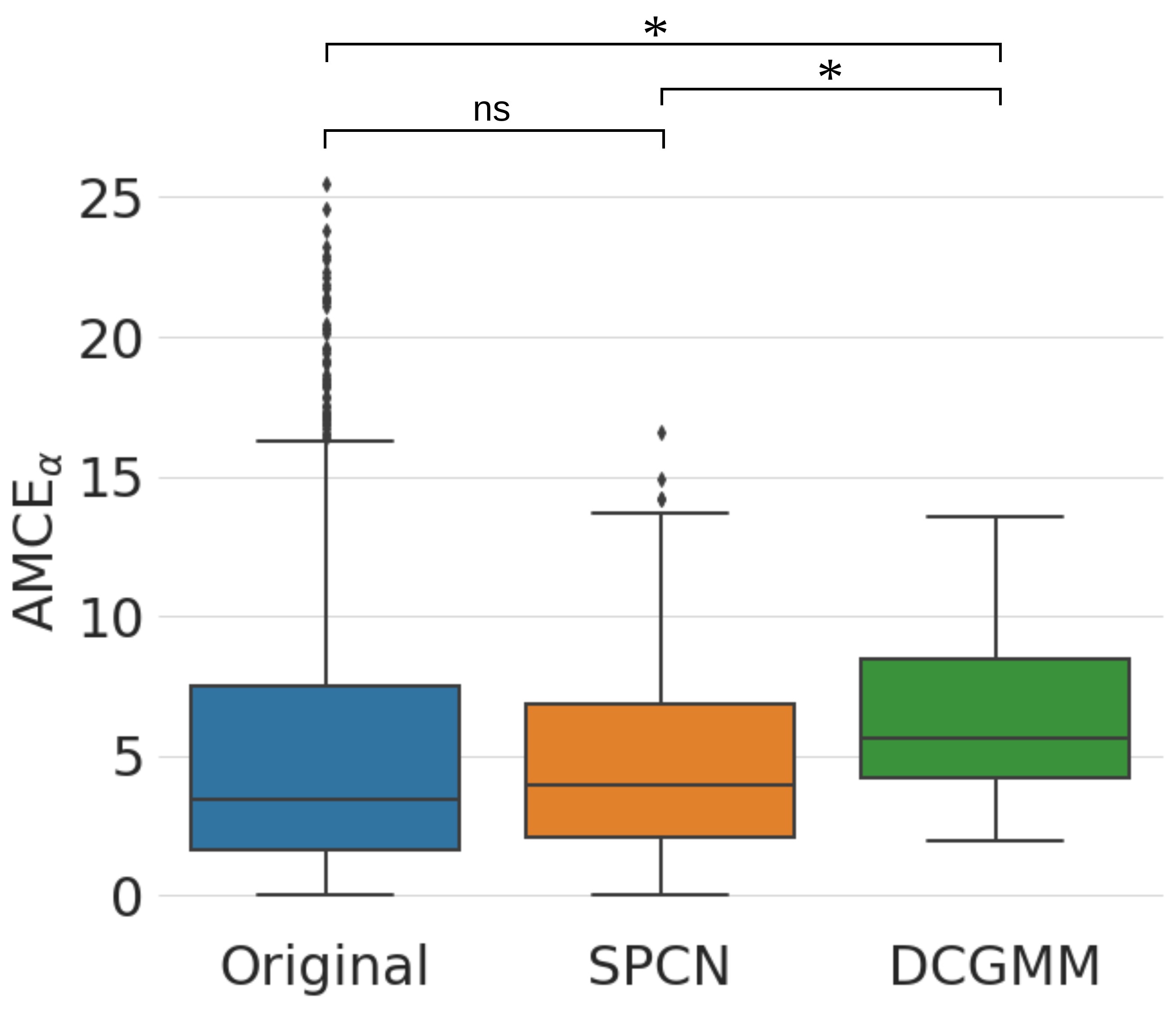}
         \label{fig:amce_a}
     \end{subfigure}
     \hfill
     \begin{subfigure}[b]{0.49\columnwidth}
         \centering
         \includegraphics[width=\textwidth]{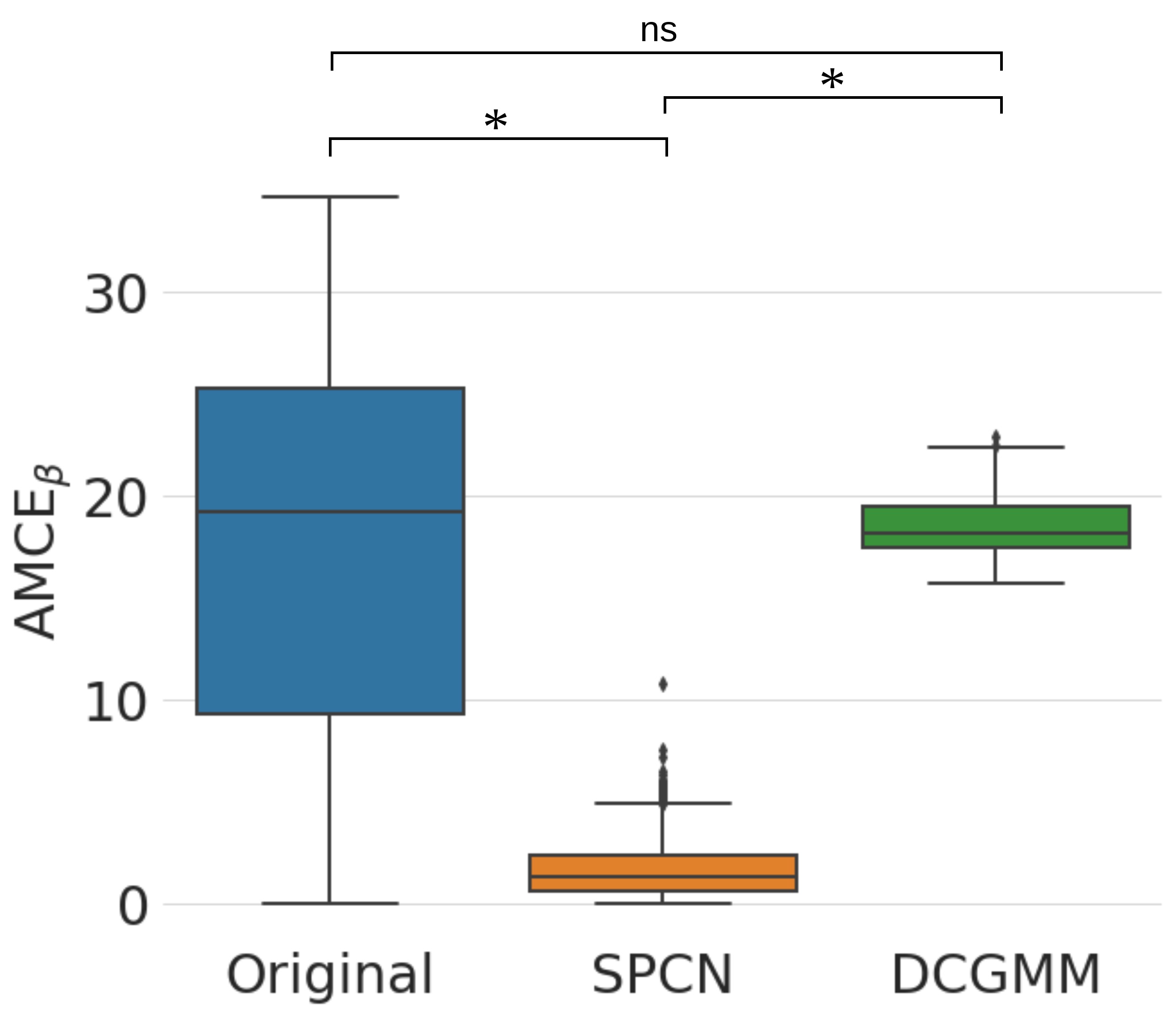}
         \label{fig:amce_b}
     \end{subfigure}

     \caption{\textbf{Box plots of NMI, NMH, AMCE$_{\alpha}$ and AMCE$_{\beta}$ values for all stain normalization methods in Experiment I.} * denotes the statistical significance of $p < 0.01$, and ns denotes not statistical significance.}
     \label{fig:sn_boxplots}
\end{figure}

Deep convolutional Gaussian mixture models (DCGMM) yielded the lowest standard deviation (SD) and coefficient of variation (CV) for both Normalized Median Intensity (NMI) and Normalized Median Hue (NMH) metrics, as indicated in \autoref{tab:exp1}. Since NMI qualifies the color consistency of the nuclei \cite{bejnordi2015stain} and NMH quantifies the global color variation of an image population \cite{pontalba2019assessing}. Thus, the results indicate that DCGMM provides qualitatively similar color distributions for nuclei with less color variation within the image population (see \autoref{fig:compare_sn}). Comparing the original and the Structure Preserving Color Normalization (SPCN), the box plots in \autoref{fig:sn_boxplots} demonstrate that DCGMM has the smallest spread of NMI and NMH values around the median (inter-quartile range) with variance statistical significance ($p < 0.01$).

Next, we evaluate the background error of each image population. DCGMM has a significantly higher mean Absolute Mean Color Error in $\alpha$ space (AMCE$_{\alpha}$) than the original image population. For $\beta$ space (AMCE$_{\beta}$), DCGMM does not have statistical significance compared to the original. It suggests that the DCGMM-processed images contain more or equal background errors than the original images, contradicting the goal of reducing color variations.

Even though SPCN does not have statistical significance with the original at AMCE$_{\alpha}$, at AMCE$_{\beta}$, SPCN provides significantly less error than the original. Moreover, both values of SPCN are statistically significantly less than DCGMM.

Therefore, we decided to implement SPCN in our framework pipeline, as it offers lower SD and CV in NMI and NMH values than the original and better AMCE values for both $\alpha$ and $\beta$ spaces than DCGMM.

\begin{table}[b]
\caption{Standard deviation and coefficient of variation for NMI and NMH, along with mean values and standard deviation for AMCE, for three different image populations.}
\label{tab:exp1}
\centering
    \begin{tabular}{@{}ccccccc@{}}
    \toprule[0.2em]
    \multicolumn{1}{c}{Method} &\multicolumn{2}{c}{NMI} &\multicolumn{2}{c}{NMH} & \multicolumn{1}{c}{AMCE$_{\alpha}$} & \multicolumn{1}{c}{AMCE$_{\beta}$}\\
    \cmidrule(r){2-3} \cmidrule(l){4-5} \cmidrule(l){6-7}
                    & SD & CV           & SD & CV           &\multicolumn{2}{c}{Mean $\pm$ SD} \\
        \midrule[0.1em]
        Original    & 0.082 & 0.137     & 0.039 & 0.042     & 5.30 $\pm$ 5.06             & 17.28 $\pm$ 9.34\\
        SPCN        & 0.057 & 0.086     & 0.012 & 0.012     & \textbf{4.67} $\pm$ 3.34    & \textbf{1.64} $\pm$ 1.33\\
        DCGMM       & \textbf{0.026} & \textbf{0.044}   & \textbf{0.008} & \textbf{0.008}   & 6.43 $\pm$ 2.69     & 18.47 $\pm$ 1.33\\
    
        \bottomrule[0.2em]\\
    \end{tabular}
\end{table}

\subsubsection{Experiment II: Backbone model evaluation}
\label{sec:result_exp2}
\begin{figure}[h!]
    \centering
    \includegraphics[width=0.8\columnwidth]{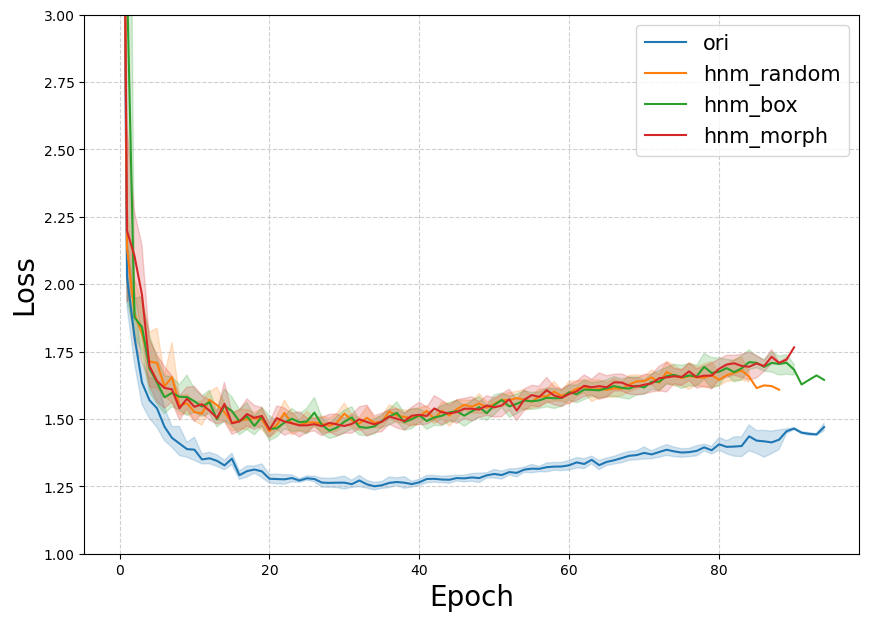}
    \caption{\textbf{Mean validation losses for each model during training.} The graph illustrates the mean validation losses calculated from \textit{class loss}, \textit{box loss}, and \textit{distributional focal loss} for each model based on the YOLOv8 architecture. The 95\% CI error bars were derived across 10-fold cross-validation.}
    \label{fig:loss}
\end{figure}

\begin{figure*}[t]
    \centering
    \includegraphics[width=0.99\textwidth]{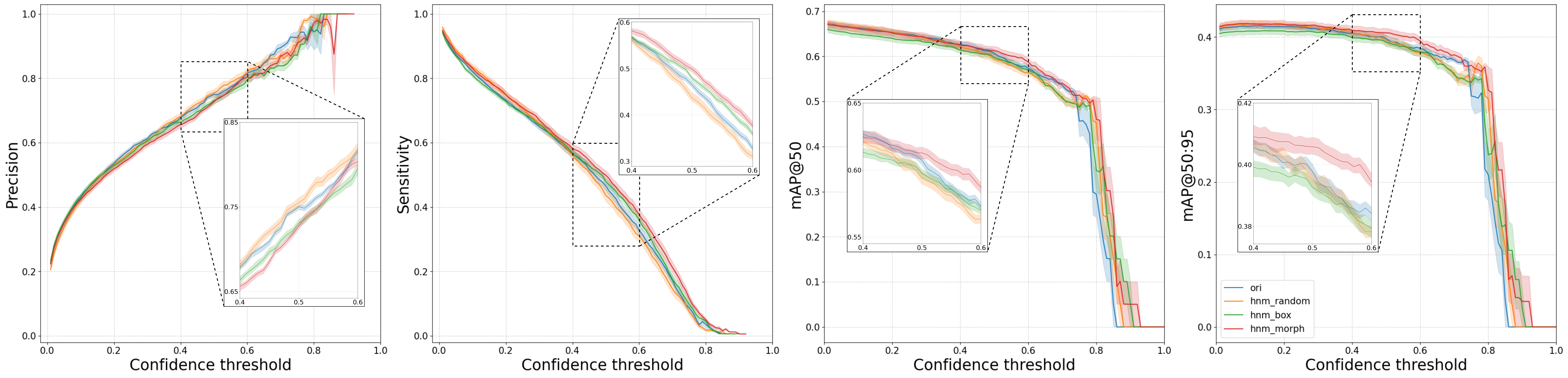}
    \caption{\textbf{Centroblast detection results:} Precision, Sensitivity, mAP at 0.5 IOU threshold (mAP@50), and mAP at 0.5 to 0.95 IOU threshold (mAP@50:95) of each model on test dataset with standard deviation error bar.}
    \label{fig:obj_det}
\end{figure*}

\begin{figure*}
    \centering
    \includegraphics[width=0.99\textwidth]{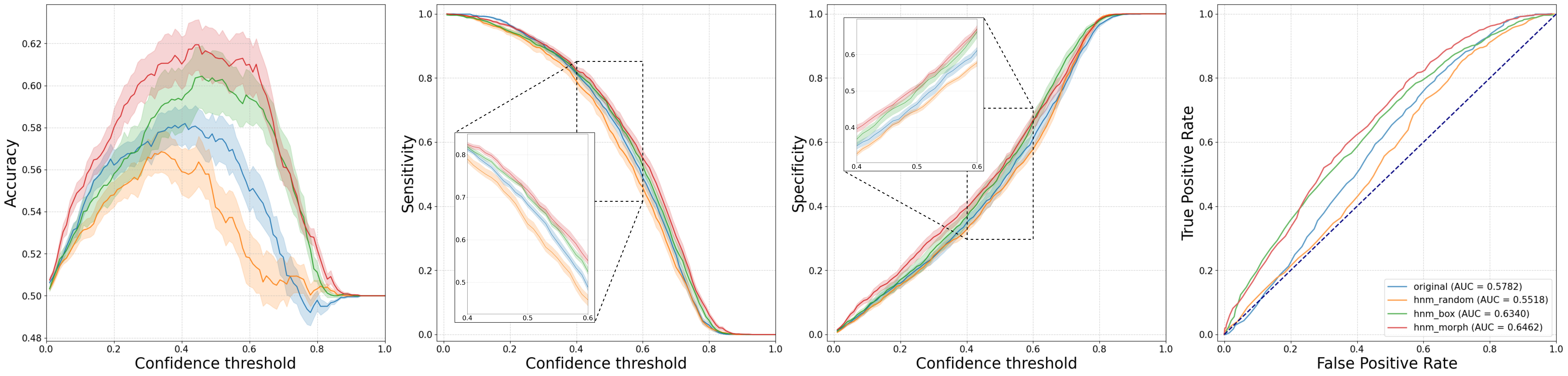}
    \caption{\textbf{Patch classification results:} Accuracy, Sensitivity, Specificity, and Receiver Operating Characteristics (ROC) curves of each model on test dataset with standard deviation error bar.}
    \label{fig:classifi}
\end{figure*}

This experiment benchmarked models trained using conventional and HNM approaches for object detection and image classification tasks. Before comparing the performance of the models, we first investigated the optimization process of each model during training. As indicated in \autoref{fig:loss}, it was observed that the validation loss of original (\textit{ori}) model reached the lowest value when converged. Other models (i.e., all HNM approaches) exhibited a similar pattern of converging more rapidly than the \textit{ori} model, albeit with a higher loss. Despite the model trained with HNM of morphological features (\textit{hnm\_morph}) going with the same trend as other HNM approaches, it achieved the highest performance in terms of mean average precision (mAP@50) and accuracy, as shown in 
\autoref{fig:obj_det} and \autoref{fig:classifi}, respectively.

For the object detection task, \autoref{fig:obj_det} provided an overview of model performance on each metric with a confidence interval. All models follow the same trend. The \textit{hnm\_morph} achieved slightly better sensitivity, but in contrast, the trade-off appears to be on precision. Nevertheless, its performance is superior to other models in mAP@50 and mAP@50:95.

For the image classification task, the \textit{hnm\_morph} outperformed all other models, especially to the model trained with random HNM (\textit{hnm\_random}). Notice that the performance of \textit{hnm\_random} dropped from \textit{ori} on all metrics. In contrast, \textit{hnm\_morph}, which was trained with the same approach but with a more reasonable undersampling method, improved its performance over the \textit{ori}. 

\begin{figure*}
    \centering
    \includegraphics[width=0.99\textwidth]{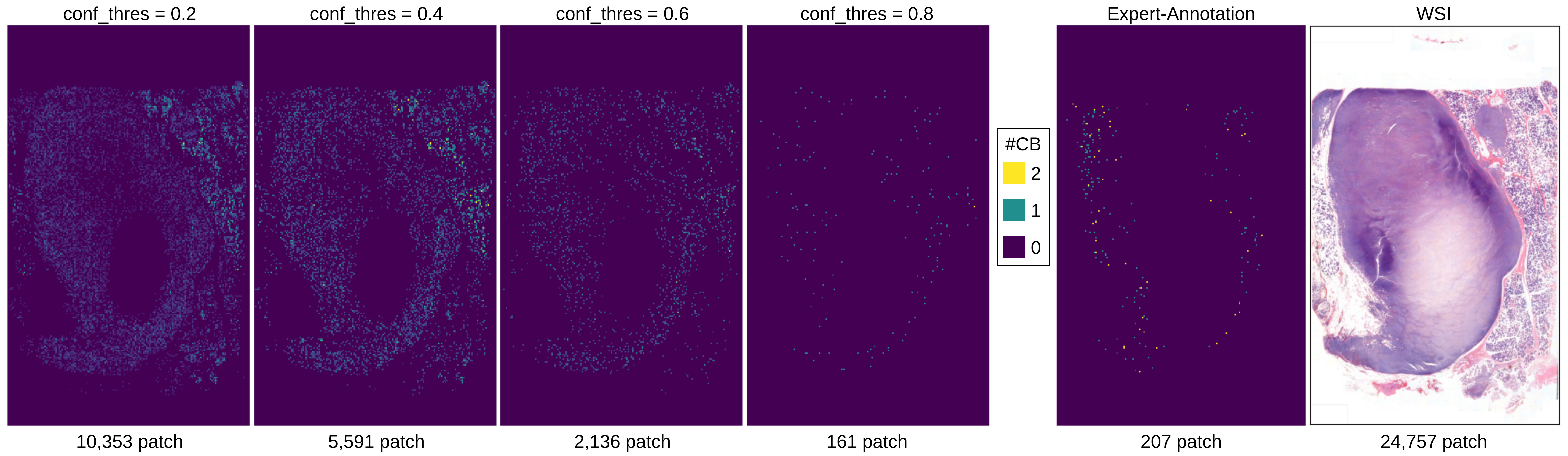}
    \caption{\textbf{Heatmap visualization of detected centroblast in each patch on an unseen WSI.} As conf\_thres increases, the bright patch becomes sparser and more prone to expert annotation. }
    \label{fig:heatmap}
\end{figure*}

\section{Discussion}
\subsection{Effect of Training with Pseudo-Negative Labels on the Model Performance}
Concerning the impact of training with pseudo-negative labels from the hard negative mining (HNM) technique over the conventional training approach, we demonstrated improvements in centroblast (CB) detection and patch classification. However, the validation loss of these models is higher than that of the conventional training method (\autoref{fig:loss}). This is because the softmax function divides the total probability mass across multiple classes (i.e., CB and non-CB) rather than just one class of CB. When computing each class's confidence score in the bounding box using the softmax function, the number of divisors is increased to two, resulting in the confidence scores for each class becoming smaller on average as the number of classes increases. Finally, the class loss of the YOLO model, which is the cross-entropy loss, in models trained with HNM is higher than in the \textit{ori} model.

\subsection{Effect of Undersampling approaches on the Model Performance}

Since our framework was designed to imitate the training loop with the pathologist's refined labels, which identifies CB based on cell color and morphology, we retrieved false-positive samples and fed them to the model. We obtain the following models based on three undersampling approaches: \textit{hnm\_random}, \textit{hnm\_box}, and \textit{hnm\_morph}. As the result of object detection and image classification tasks, the model with the morphological features (\textit{hnm\_morph}) performs best. Suggest that the design of the undersampling approach is essential to take advantage of the HNM technique and that pathologists' intuition still provides some information for deep learning in order to distinguish between non-CB cells and actual CB cells.

\subsection{Guidance for Clinical Implementation and Future Works}
In conventional histopathological workflow, pathologists count the number of centroblasts (CB) in ten randomly selected high-power fields (HPF), leading to high inter- and intra-observer variability and being vulnerable to sampling bias. The inter- and intra-observer variability among pathologists is crucial since it directly impacts patient grading and management \cite{lozanski2013inter}. In order to reduce the variability between pathologists, a solid guideline for finding potential HPF in WSI is one solution.

With \textit{PseudoCell} framework, pathologists will obtain two guidelines (i) heatmap visualization for potential CB regions in WSI and (ii)  CB annotations at HPF level for identifying CB. Pathologists' remaining job is to select the HPF and then accept the annotation or self-identify CB cells. Pathologists are only required to set the confidence threshold (conf\_thres), which can range from zero to one, when using \textit{PseudoCell}. The conf\_thres parameter determines the initial confidence level of CB annotations reported to pathologists. A low conf\_thres (conf\_thres = 0.2) produces a dense heatmap in \autoref{fig:heatmap}, whereas a high conf\_thres (conf\_thres = 0.8) produces a sparse heatmap that more closely resembles the expert-annotation.

We will divide the histopathological process into HPF selection phase and CB identification phase. In each phase, the real-world adjustment of conf\_thres to facilitate pathologist preference could take the form of the following suggestion:

\subsubsection{HPF selection phase} with a high conf\_thres, \textit{PseudoCell} offers a sparse HPF that is still sufficient to grade FL, which is suitable for pathologists who wish to complete the grading task rapidly. In contrast, when conf\_thres is low, \textit{PseudoCell} generates a dense heatmap that identifies the region containing intensive CB and regions with less CB. This approach is suitable for pathologists who wish to determine the HPF independently.

\subsubsection{CB identification phase} A high conf\_thres is advantageous for pathologists who prefer to self-identify on CB with some framework-suggested CB annotation. In contrast, a low conf\_thres will enable the framework to recommend more CB annotation, which is ideal for pathologists who wish to check off the annotation. 

The pathologists' workload can be reduced by \textit{PseudoCell} accurately narrowing down the areas requiring their attention during examining tissue as in \autoref{fig:heatmap}. From all 24,757 patches with tissue in WSI, the framework highlights 10,353 and 161 patches that contain potential CB candidates based on conf\_thres with an inference time of approximately 0.03 seconds per patch. In other words, the framework can eliminate 58.18 to 99.35\% of all WSI patches that do not appear to be CB candidates at the conf\_thres. Pathologists can therefore focus on identifying CB on the slide. We anticipated that inter- and intra-variability of pathologists would decrease after implementing our framework in the real world. In contrast, the machine benefits from pathologists providing actual CB as refined labels. These labels can be used to improve the model's performance in the future. This cycle leads to human-machine collaboration in the real world, which is one of the objectives of this work.

\textit{Pseudocell} can also provide a second opinion that offers additional safety for patients and instills greater confidence in doctors, enhancing their efficiency and reducing the likelihood of errors. For instance, when there is a need to distinguish between an infection and follicular cell lymphoma, particularly in its early stages, Thai pathologists, who may already be handling a heavy workload, could potentially issue a false negative, especially when the pathological area is small.

Integrating \textit{PseudoCell} into the histopathological workflow offers several benefits. Firstly, the model assists pathologists by highlighting regions or suggesting potential CB cell candidates within the tissue, thereby narrowing the examination focus. It serves as an additional quality control mechanism, flagging areas that may contain CB cells and assisting pathologists in not overlooking significant findings, thereby reducing diagnostic errors. In addition, pathologists can use the model's predictions as a benchmark to compare and contrast their observations. This iterative process improves their skills in order to recognize centroblast cells, thereby enhancing diagnostic precision over time. Incorporating \textit{PseudoCell} contributes to improved efficiency, quality control, and training and education for identifying centroblast cells in histopathology.

Lastly, the implementation of the \textit{PseudoCell} framework indeed represents a cost-effective investment. When compared to the recurrent expenses associated with sustaining a team of pathologists – including salaries, benefits, and training costs – the financial outlay required to incorporate and maintain this AI model is comparatively minimal. Moreover, the application of this technology can drastically boost efficiency by expediting the process of identifying centroblasts, thereby allowing pathologists to concentrate on more intricate tasks. This surge in efficiency can, in turn, decrease operational costs over time, as a speedier diagnosis may result in reduced lab usage and quicker patient turnover.

In future work, if there are additional object detection models or updated versions, the \textit{PseudoCell} framework permits their implementation by modifying the backbone model. Furthermore, dealing with data limitations and model transparency is crucial for pathologists to understand and have confidence in the model's decision-making. Combining weakly supervised paradigm (e.g., MIL or Attention Map) with explainability techniques (e.g., LIME, SHAP, and CAM) is a promising next step to investigate.

\section{Conclusion}
In conclusion, our study introduces the PseudoCell framework for centroblast (CB) cell detection, which enhances the performance of the backbone model by using false-positive samples from the Hard Negative Mining (HNM) method as pseudo-negative labels. PseudoCell effectively distinguishes between actual CB and non-CB cells in patches from whole-slide images (WSI). Our experiments and evaluations demonstrate that model training from HNM on Neighborhood-based Recursive search undersampling using morphological features achieves the best results in CB detection and patch classification tasks. PseudoCell can reduce pathologists’ workload by accurately identifying tissue areas requiring attention during examination. Depending on the confidence threshold, PseudoCell can eliminate 58.18-99.35\% of non-CB tissue areas on WSI. Furthermore, PseudoCell can serve as a second opinion to differentiate between infection and follicular cell lymphoma, particularly in the early stages, making it cost-efficient for quality control and educational purposes in CB recognition. This study presents a practical centroblast prescreening method that does not rely on pathologists’ refined labels for improvement. It suggests the potential for human-machine collaboration in CB identification, alleviating the burden on clinicians by focusing their labeling efforts on regions suggested by PseudoCell, rather than manual labeling as conventionally done.

\section*{Acknowledgment}
The authors wish to thank Surat Phumphuang for her coordination in operating the research.


\bibliographystyle{IEEEtran}
\bibliography{ref}

\end{document}